# Uso del software tracker para la enseñanza de los laboratorios de física a distancia: Demostración de la disminución de error en el péndulo simple

# Use of Tracker software for teaching distance physics laboratories: Demonstration of error reduction in the simple pendulum

Ariatna Isabel Zamalloa Ponce de Leon
Universidad Nacional de San Antonio Abad del Cusco. Cusco, Perú
182773@unsaac.edu.pe
https://orcid.org/0000-0002-0627-9896

Gerson Tito Cruz Sulla
Universidad Nacional de San Antonio Abad del Cusco. Cusco, Perú
184159@unsaac.edu.pe
https://orcid.org/0009-0004-2442-3507

Juan Carlos Camani Ccollana
Universidad Nacional de San Antonio Abad del Cusco. Cusco, Perú
193803@unsaac.edu.pe
https://orcid.org/0009-0007-0145-2446

Valery Samara Almanza Zuñiga
Universidad Nacional de San Antonio Abad del Cusco. Cusco, Perú
192962@unsaac.edu.pe
https://orcid.org/0009-0003-5311-8432

Brayan Gonzales Huisa
Universidad Nacional de San Antonio Abad del Cusco. Cusco, Perú
192965@unsaac.edu.pe
https://orcid.org/0009-0000-9228-3359

Diego Reiner Villa Recharte
Universidad Nacional de San Antonio Abad del Cusco. Cusco, Perú
182771@unsaac.edu.pe
https://orcid.org/0000-0002-2999-1237

Bruce Stephen Warthon Olarte
Universidad Nacional de San Antonio Abad del Cusco. Cusco, Perú
160531@unsaac.edu.pe
https://orcid.org/0000-0002-0985-1579

**Resumen**

Este estudio se centra en la evaluación de tres métodos utilizados para determinar la gravedad teórica en la ciudad del Cusco y su relevancia en la educación a distancia. El primer método se basa en la ley de la gravitación universal de Newton, obteniendo un valor teórico de la aceleración de la gravedad de 9.7836 m/s². El segundo método emplea un péndulo simple, que proporciona una estimación cercana al valor teórico. El tercer método utiliza el software Tracker para analizar el movimiento del péndulo, ofreciendo una mayor precisión en las mediciones.

La importancia de esta investigación en la educación a distancia radica en la necesidad de herramientas y estrategias efectivas para la enseñanza de asignaturas prácticas como la Física. Tanto el péndulo simple como el software Tracker brindan soluciones viables para realizar experimentos y análisis de datos de manera remota. Estas herramientas permiten a los estudiantes adquirir conocimientos teóricos y prácticos, desarrollar habilidades de medición y análisis de datos para promover un aprendizaje más significativo.

**Palabras clave:** gravedad teórica, educación a distancia, péndulo simple, software Tracker.

**Abstract**

This study focuses on evaluating three methods used to determine the theoretical gravity in the city of Cusco and their importance in distance education. The first method is based on Newton's law of universal gravitation, obtaining a theoretical gravity value of 9.7836 m/s². The second method employs a simple pendulum, which provides an estimation close to the theoretical value. The third method uses the Tracker software to analyze the pendulum's motion, offering higher precision in measurements.

The significance of this research in distance education lies in the need for effective tools and strategies to teach practical subjects like physics. Both the simple pendulum and Tracker software provide viable solutions for conducting experiments and data analysis remotely. These tools allow students to acquire theoretical and practical knowledge, develop measurement and data analysis skills to promote more meaningful learning.

## 1. Introducción

La Ciencia y la Física, como parte de esta, brinda al ser humano distintos conocimientos tanto teóricos como prácticos, es fundamental para la sociedad que muchos conceptos teóricos puedan ser enseñados al público en general, sea acoplada la realización de experimentos para un mayor entendimiento de la teoría fundamental (Franklin y Perovic, 2023). Esta idea se refleja en la enseñanza de la misma, ya sea en la educación secundaria o universitaria. Dentro de los cursos de física es necesario desarrollar diferentes experimentos con la finalidad de complementar y ampliar los conocimientos previos, así como, adquirir habilidades en el saber, conocer y hacer general. (Massoni y Moreira, 2010; Guisasola et al., 2012). El desarrollar conceptos de física a través del trabajo experimental en los laboratorios, nos lleva a la adquisición de aprendizaje significativo, que se da cada vez que el estudiante utiliza sus conocimientos básicos previos y, apoyados en el uso de herramientas prácticas en el laboratorio y el análisis de datos, facilita a que el campo conceptual se estructure y enriquezca (Agudelo y García, 2010).

Desafortunadamente en muchos países, la necesidad de laboratorios de enseñanza que estén debidamente equipados o con un inventario actualizado, a menudo limita o dificulta la realización de actividades prácticas. Adicionalmente, se presentan problemas con respecto al elevado número de alumnos, lo cual está directamente relacionado con la falta de mayor cantidad de laboratorios o el mismo espacio para la enseñanza de la ciencia (Anni, 2021). En términos generales, el gasto relacionado con la adquisición de todo tipo de materiales requeridos para llevar a cabo numerosas prácticas de laboratorio tiende a ser sumamente alto, este costo se incrementa aún más si se desea contar con múltiples ejemplares de diferentes prácticas, de manera que, la mayoría de estudiantes pueden llevar a cabo sólo un número determinado de experimentos (Rebollo y González, 2016).

Recientemente debido a la situación de confinamiento ocasionada por la pandemia de COVID-19, según un artículo de la Organización Mundial de la Salud (Inmunidad colectiva, confinamientos y COVID-19, s/f), el uso de los recursos cotidianos que el estudiante tenía para realizar sus avances prácticos de manera física, como es el uso de los laboratorios y su equipamiento, se vio restringido por el distanciamiento físico. En ese sentido se optó por el uso de recursos de aprendizaje en línea de alta calidad y eficiencia, que permitió a estudiantes de diversos niveles educativos satisfacer y reemplazar sus necesidades de estudio y aprendizaje, lo cual a su vez les posibilita realizar trabajos de laboratorio e investigación de manera virtual e híbrida (Ray y Srivastava, 2020).

Por todo lo expuesto, es importante resaltar que el desarrollo de nuevos métodos de aprendizaje experimental sustituyen el uso de los laboratorios convencionales (Anni, 2021). Esto es posible poner en práctica gracias a la implementación de herramientas digitales mediante el uso de computadoras y simuladores, estos mismos ayudarían a que los estudiantes desarrollen hábitos, destrezas y habilidades mentales analíticas para mejorar la formulación de soluciones a problemas prácticos; ya que los simuladores son útiles para reducir el tiempo de respuestas de los participantes y facilita su comprensión más sencilla y eficiente de fenómenos mundiales (Domínguez, 2016). Como ejemplo tenemos que el análisis de video o el estudio de estos fotogramas en la educación y sobre todo en la física recibe mucha atención porque en base a la visualización se puede hacer que las lecciones tradicionales sean más interesantes y más efectivas durante el aprendizaje remoto o híbrido. El análisis correspondiente al video posterior a la prueba destaca este método como uno de los métodos más innovadores para la enseñanza de la física, lo que hace que la ciencia sea más interesante y llamativa que nunca para cualquier estudiante de cualquier nivel de estudios (Ramli et al., 2016).

Algunos de estos experimentos en laboratorios para la enseñanza de la física que se realizan comúnmente son los aplicados a determinar la aceleración de la gravedad en un lugar, tales como caída libre, péndulo simple, entre otros, (Mendoza, 2018). Por ese motivo, para el presente trabajo es necesario conocer la determinación de la aceleración de la gravedad mediante el péndulo simple usando dos métodos, uno el método convencional y otro mediante el software Tracker, donde también es necesario evaluar el error porcentual de estos métodos en función a la gravedad teórica, para de esta manera demostrar que método es más preciso e idóneo (Rodríguez, 2020).

**1.1. Determinación de la gravedad teóricamente**

Una de las muchas maneras para determinar la gravedad teórica es usando la ley de la gravitación universal de Newton, la cual matemáticamente nos permite determinar la fuerza de interacción a grandes distancias de dos cuerpos (Sebastiá, 2013). Donde la ecuación representativa de la ley de gravitación universal es:

$$F_R = G \cdot \frac{m \cdot M_T}{R^2} \tag{1}$$

Donde:
$F_R$ : Fuerza Resultante.
$G$ : Constante Gravitacional cuyo valor es $6,67x10^{-11}$ N.m²/kg² (CODATA, 2023).
$m$ : Masa de un cuerpo en interacción con la tierra.
$M_T$ : Masa de la tierra cuyo valor es $5.96x10^{24}$ kg (NASA, 2023).
$R$ : Radio del centro de la Tierra en función a la ciudad de Lima cuyo valor es $6,371x10^6$ m (WIKIWAND, 2023).

**1.2. Péndulo simple**

Un péndulo simple es un modelo ideal de un sistema mecánico que muestra un movimiento periódico, se compone por una masa $m$ suspendida de una cuerda delgada con masa despreciable y de longitud $L$, el movimiento se produce en un plano vertical y es impulsado por la fuerza de la gravedad, el movimiento será muy parecido al de un movimiento armónico simple si el ángulo $\theta$ es menor a 10° (Serway y Jewett, 2009; Hugh y Freedman, 2009; José Martínez, 2015).

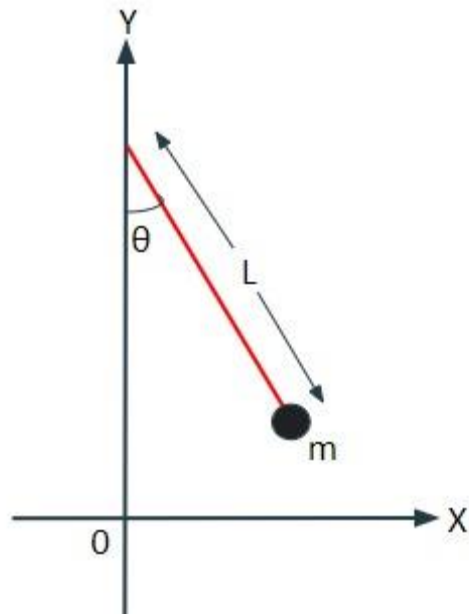

*Figura 1. Descripción gráfica del péndulo simple*

A partir del análisis de la Figura 1, la ecuación diferencial para el Movimiento Armónico Simple (MAS) es:

$$\frac{d^2\theta}{dt^2} + \omega^2 \sin\theta = 0 \qquad (2)$$

Para ángulos menores a 10° la ecuación (2) toma la forma de:

$$\frac{d^2\theta}{dt^2} = -\frac{g}{L}\theta \qquad (3)$$

Donde:

$$\omega = \sqrt{\frac{g}{L}} \qquad (4)$$

La ecuación (3) es una ecuación diferencial homogénea de segundo orden con coeficientes constantes. Su solución general se expresa mediante exponenciales complejos (Solís-Rivera et al., 2023), aplicando las condiciones iniciales cuando el valor de $\theta(t=0) = 0$, se obtiene:

$$\theta(t) = A\sin(\omega t + \phi) \qquad (5)$$

Donde: $A$ es la amplitud, $\phi$ es el ángulo de desfase y la expresión análoga de la frecuencia angular es:

$$\omega = 2\pi f \qquad (6)$$

Como la frecuencia es inversamente proporcional al periodo, es decir $f = 1/T$; tenemos:

$$\omega = \frac{2\pi}{T} \qquad (7)$$

Reemplazando la ecuación (7) en la ecuación (4) y despejando $T$, el periodo del movimiento sería:

$$T = 2\pi\sqrt{\frac{L}{g}} \qquad (8)$$

Despejando $g$ tenemos que:

$$\frac{T}{2\pi} = \sqrt{\frac{L}{g}} \;\Rightarrow\; \frac{T^2}{4\pi^2} = \frac{L}{g}$$

Resultando que el valor de la aceleración de gravedad sería:

$$g = \frac{4\pi^2 L}{T^2} \quad (9)$$

Esta es la ecuación con la que se puede calcular el valor de la aceleración de la gravedad usando datos experimentales del péndulo simple.

Sin embargo, otra manera de evaluar este valor, sería en función a la ecuación (4), donde el valor de $g$ sería:

$$g = L\omega^2 \quad (10)$$

El cual será usado para determinar el valor de la aceleración de gravedad mediante el software Tracker.

### 1.3. Software Tracker

Tracker es un software gratuito y multiplataforma de análisis de video y modelado de objetos en movimiento, el programa permite generar datos de posición, velocidad, aceleración, etc., de los objetos a ser estudiados, de esta manera con Tracker se puede adquirir bastante información partiendo de estos videos, como gráficos, marcos de referencia, puntos de calibración, perfiles de línea para el análisis de los patrones de espectros y de interferencia, así como modelos de partículas dinámicas. Usar Tracker para el análisis experimental durante el desarrollo de prácticas de laboratorios nos permite reducir los errores sistemáticos: error instrumental, error personal, error de elección de método, además de los errores accidentales, es decir las pequeñas variaciones entre las mediciones sucesivas realizadas por un mismo observador (Domínguez, 2016).

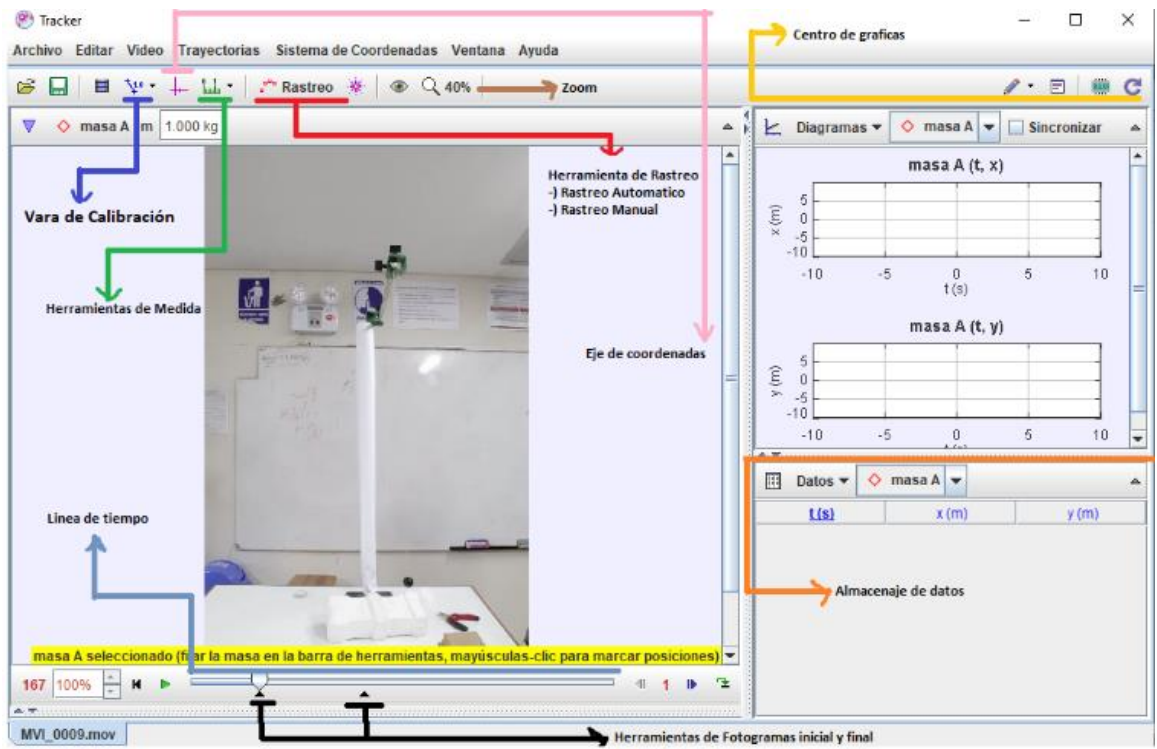

*Figura 2. Herramientas del entorno de trabajo Tracker*

Este programa está diseñado para ser utilizado en laboratorios, clases demostrativas y conferencias de ciencias, en específico la física, en las universidades y todos los niveles educativos (Garzón y Villate, 2022; Domínguez, 2016). Como resultado de todo lo expuesto previamente, el uso de herramientas científicas y tecnológicas como Tracker ha adquirido una creciente relevancia en diferentes países (King et al., 2014). En particular, la aplicación de Tracker al péndulo simple lo convierte en un laboratorio híbrido que permite a los estudiantes utilizar el software tanto físicamente durante la experimentación como virtualmente en el procesamiento y análisis de datos. Estudios previos han señalado que no existe una diferencia significativa en los resultados de aprendizaje entre entornos virtuales e híbridos (Srivastava et al., 2013).

Visto específicamente en el péndulo simple, el uso de este software ofrece diversos beneficios específicos, por ejemplo, proporciona a los estudiantes una experiencia práctica y visual en el análisis de datos, lo que contribuye a una mejor comprensión de los conceptos teóricos relacionados con el péndulo simple. Además, Tracker permite un análisis más preciso y detallado de los movimientos del péndulo, mejorando la comprensión de los estudiantes sobre este fenómeno físico (Brown, 2023).

**1.4. Python**

Python es un lenguaje de programación utilizado ampliamente en el ámbito científico y en el análisis de datos a nivel general (Herrera, 2023). En esta investigación y más específicamente al momento de usar el Tracker para el gráfico y la obtención de parámetros se puede usar por conveniencia Python, donde importamos diversas bibliotecas para facilitar nuestro análisis, entre ellas están: NumPy que nos permitió realizar cálculos numéricos y procesar datos eficientemente (Harris et al., 2020). SymPy nos ayudó con operaciones algebraicas y simplificación de expresiones matemáticas (Meurer et al., 2017). Matplotlib nos permitió generar gráficos en 2D para visualizar nuestros resultados (Hunter, 2007). Utilizamos la función curve fit de la sub biblioteca Scipy. Optimize para ajustar curvas a nuestros datos (Virtanen et al., 2020). Pandas fue esencial para leer y manipular datos almacenados que pueden ser en formato CSV o Excel (McKinney, 2010). Finalmente, Seaborn nos brindó herramientas para crear visualizaciones estadísticas de manera atractiva y comprensible (Waskom, 2021).

**1.5. Error porcentual**

Es necesario que, para la comparativa de experimentos usados, el estudiante tenga que realizar la validez de su experiencia mediante el cálculo del error porcentual, por eso para determinar el error porcentual de cada método en función al valor teórico (Herrera, L., Torres G., 2022), es necesario emplear la siguiente fórmula:

$$E\% = \frac{|V_t - V_e|}{V_e} \times 100\% \qquad (11)$$

Donde:

$V_t$: Valor teórico
$V_e$: Valor experimental

En este sentido, este estudio examinará los beneficios y aplicaciones prácticas de la herramienta Tracker en el contexto del estudio del péndulo simple. Se presentará un experimento realizado en las instalaciones de la facultad de física de la Universidad Nacional de San Antonio Abad del Cusco y se desarrollará un análisis numérico y visual utilizando el software Tracker el cual se sincronizará y escalará automáticamente con el video para una comparación directa con el mundo real. Al aprovechar las ventajas que ofrece Tracker, se espera que esta investigación contribuya al avance en la comprensión y análisis del péndulo simple en el ámbito de la investigación y la educación científica.

**2. Método**

**2.1. Determinación de la gravedad teórica:**

Primeramente, es necesario evaluar un valor teórico de la gravedad sobre el cual sustentamos la eficacia del experimento del péndulo simple y del software Tracker. Usaremos el valor teórico de la gravedad en referencia a la ciudad del Cusco, la cual hallaremos en función a la ecuación (1), para esto realizamos la adición del valor de la altitud de la ciudad del Cusco al valor de R, siendo 3 350 metros sobre el nivel del mar considerando los datos brindados por Google Earth en la facultad de Física de la Universidad Nacional de San Antonio Abad del Cusco, resulta un total de $\mathbf{6,37435 \times 10^6 m}$.

Sustituyendo en la ecuación (1) tenemos:

$$m.g = \mathbf{6,67}x\mathbf{10}^{-11}\frac{\text{N.m}^2}{\text{kg}^2} \cdot \frac{m(\mathbf{5.96 \times 10^{24} kg})}{(\mathbf{6,37435 \times 10^6 m})^2}$$

$g$ : aceleración de la gravedad de la Tierra.
Resolviendo, tenemos:

$$g \approx 9{,}7836 \text{ m/s}^2 \qquad (12)$$

### 2.2. Determinación de la gravedad mediante el uso del péndulo simple:

**Procedimiento**
Realizamos el experimento del péndulo simple usando a la vez una cámara filmadora que nos permitirá usar el mismo experimento con el mismo lanzamiento para el software Tracker.

**Construcción del péndulo simple:**
Se realiza el armado del péndulo simple de acuerdo al concepto general, usando un soporte universal que sujetada por una mordaza tiene consigo una varilla de aluminio, sobre la que sostendremos una cuerda con masa despreciable a una canica.

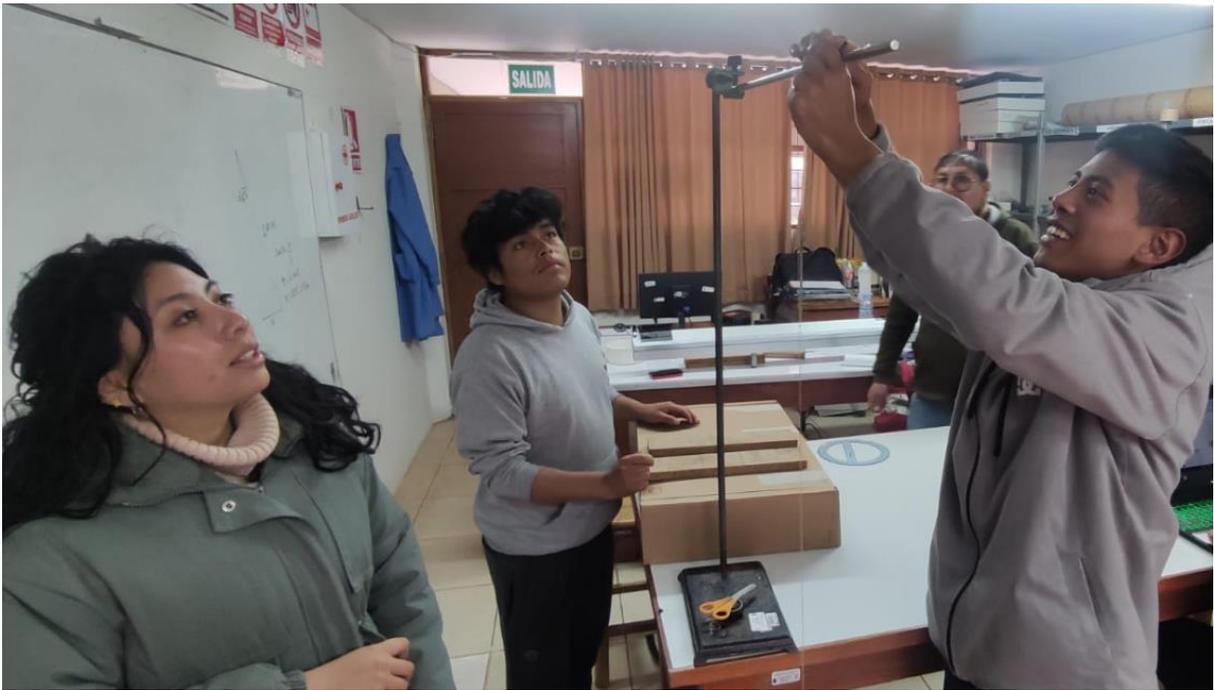

***Figura 3***. *Armado del péndulo simple*

Variamos la longitud de la cuerda de 1 m y mediremos el tiempo para 10 oscilaciones, teniendo en cuenta que el ángulo debe ser menor a 10°, para lo cual emplearemos en la ecuación:

$$x = L \sin(5°) \qquad (13)$$

De esta manera evitamos el uso de un transportador y nos permitiremos medir el ángulo en base a una recta trazada en la base en función a la ecuación (13) utilizando directamente la regla.

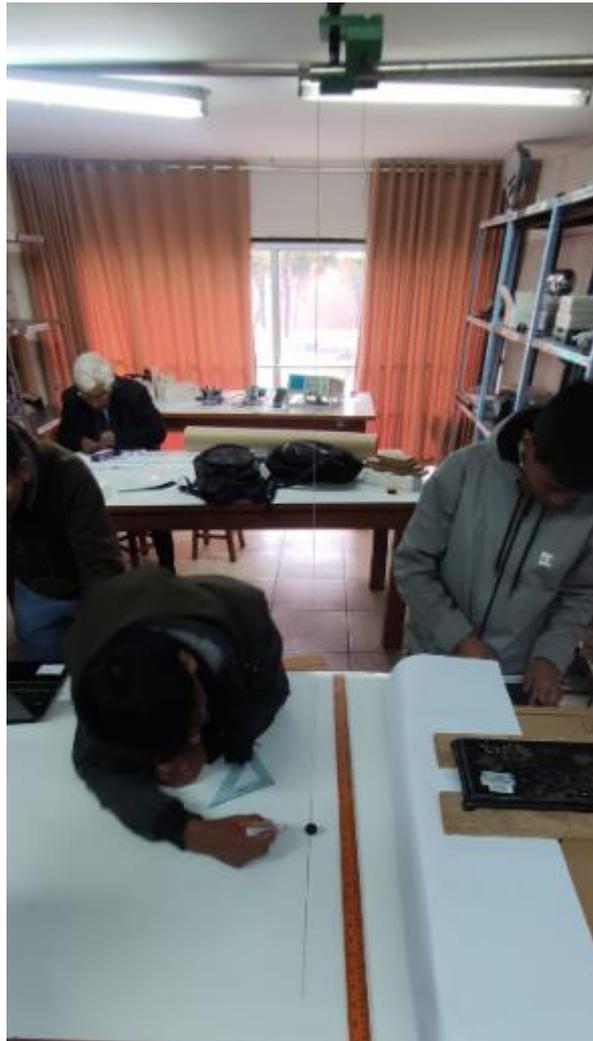

*Figura 4. Medición de la distancia de desplazamiento de la canica sobre el eje x*

Después repetimos esta misma acción disminuyendo en 5 cm la longitud de la cuerda hasta llegar a 50 cm, y llenamos los datos obtenidos en la tabla (1).

*Tabla 1.*
*Tiempos obtenidos para 10 oscilaciones del péndulo simple*

| N° de oscilaciones | L(m) | t1 | t2 | t3 | t4 | t5 | x |
|---|---|---|---|---|---|---|---|
| 10 | 1 | 19.55 | 19.64 | 19.55 | 19.63 | 19.70 | 0.087 |
| 10 | 0.95 | 19.35 | 19.30 | 19.24 | 19.18 | 19.14 | 0.083 |
| 10 | 0.9 | 18.93 | 18.88 | 18.60 | 18.56 | 18.75 | 0.078 |
| 10 | 0.85 | 18.15 | 18.45 | 18.25 | 18.40 | 18.40 | 0.074 |
| 10 | 0.8 | 17.73 | 17.58 | 17.45 | 17.41 | 17.42 | 0.07 |

| 10 | 0.75 | 17.23 | 17.20 | 17.01 | 17.43 | 17.38 | 0.065 |
| 10 | 0.7  | 16.33 | 16.35 | 16.37 | 16.43 | 16.34 | 0.061 |
| 10 | 0.65 | 15.79 | 15.73 | 15.69 | 15.66 | 15.78 | 0.057 |
| 10 | 0.6  | 15.20 | 15.19 | 15.45 | 15.50 | 15.60 | 0.052 |
| 10 | 0.55 | 14.78 | 14.71 | 14.47 | 14.78 | 14.95 | 0.048 |
| 10 | 0.5  | 13.77 | 13.54 | 13.88 | 13.88 | 13.30 | 0.044 |

Fuente: Elaboración propia.

El periodo se obtuvo mediante la ecuación:

$$T = \frac{tiempo\ medido}{n\acute{u}mero\ de\ oscilaciones}$$

Posteriormente evaluaremos el valor de la aceleración de la gravedad en función a los datos obtenidos sustituyendo cada uno de estos en la ecuación (9), de tal manera obtenemos la tabla (2).

*Tabla 2.*
*Valores de la aceleración de la gravedad en función de los periodos calculados*

| $n$ | $t$ (s) tiempo | $T$(s) periodo | $T^2$ | $L$ (m) | $g$ (m/s^2) |
|---|---|---|---|---|---|
| 1  | 19.6140 | 1.9614 | 2.0010 | 1.0000 | 10.2619 |
| 2  | 19.2450 | 1.9245 | 1.9657 | 0.9500 | 10.1262 |
| 3  | 18.8400 | 1.8840 | 1.8997 | 0.9000 | 10.0101 |
| 4  | 18.2750 | 1.8275 | 1.8600 | 0.8500 | 10.0476 |
| 5  | 17.5180 | 1.7518 | 1.8047 | 0.8000 | 10.2915 |
| 6  | 17.2500 | 1.7250 | 1.7287 | 0.7500 | 9.9505 |
| 7  | 16.3640 | 1.6364 | 1.6727 | 0.7000 | 10.3200 |
| 8  | 15.7300 | 1.5730 | 1.6173 | 0.6500 | 10.3709 |
| 9  | 15.3880 | 1.5388 | 1.5563 | 0.6000 | 10.0034 |
| 10 | 14.7380 | 1.4738 | 1.4923 | 0.5500 | 9.9964 |
| 11 | 13.6740 | 1.3674 | 1.4387 | 0.5000 | 10.5570 |
|    |         |        |        | $\bar{g}$ | 10.1760 |

Fuente: Elaboración propia.

Donde encontramos que el promedio de la gravedad evaluada según cada variación de longitud de la cuerda en el péndulo simple fue de 10.1760 m/s², con una incertidumbre de 0.1961 m/s²

## 2.3. Determinación de la gravedad mediante el uso del software Tracker aplicado al péndulo simple:

**Procedimiento**

1. Se prepara un ambiente adecuado para el set de grabación asegurándonos que el objeto de estudio resalte en el escenario para garantizar que la herramienta de rastreo automático no presente problemas.
2. Verificamos que los fotogramas de la cámara estén comprendidas entre 60-120 fotogramas por segundo, esto ayudará a que el objeto de estudio no se deforme en la grabación.
3. Posicionamos la cámara perpendicularmente al plano de estudio y paralelo al objeto a seguir para fijar que el objeto esté enfocado.
4. Una vez grabado el experimento del péndulo simple, convertimos el formato del vídeo a MOV.
5. Luego en el software, importamos el video correspondiente, seleccionamos los fotogramas iniciales y finales donde empezamos con el estudio.
6. Usando la vara de calibración (Figura 2) medimos las referencias métricas del objeto de estudio, previamente a esto conocemos la medida real de estas referencias, lo cual ayudará a que los datos obtenidos sean consistentes con la realidad.
7. Posicionamos los ejes de coordenadas en el centro de masa del objeto de estudio.
8. Seleccionamos la herramienta de rastreo, luego agregamos un nuevo rastreo y seleccionamos masa puntual, presionamos shift + control y seleccionamos el objeto de estudio.
9. Se abrirá una ventana del auto rastreador, en el video debemos asegurar que el objeto esté completamente seleccionado, el rastreador automático funcionará correctamente haciendo clic en "Buscar".
10. Una vez terminado el rastreo, Tracker generará una base de datos los cuales estarán en función de la posición y el tiempo como también incluirá los gráficos correspondientes como se muestra en la Figura 6 y Figura 7.
11. Exportamos los datos presionando el clic derecho, seguidamente en la opción de copiar datos seleccionados, pulsamos la opción como formateado.
12. Los datos pueden ser exportados a cualquier programa de procesamiento de datos como Excel, SPSS, Python, etc.

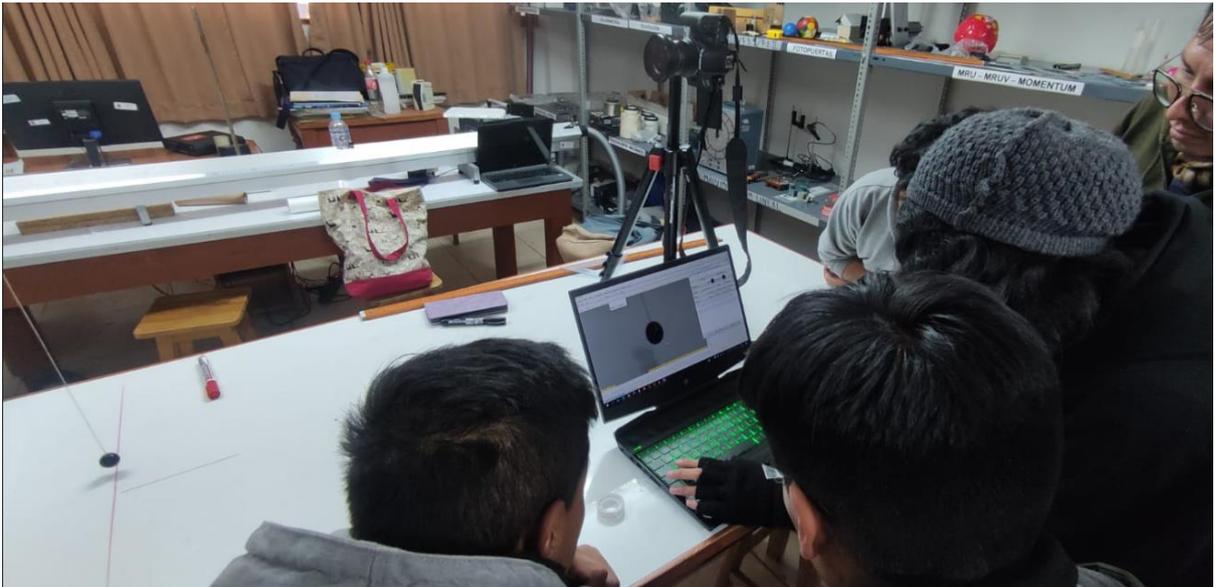

*Figura 5.* *Uso del software Tracker para el análisis del péndulo simple.*

**Recolección de datos**

Para el cálculo de la aceleración de la gravedad realizamos una serie de 11 vídeos en función al desplazamiento de la canica, de los cuales se obtienen los siguientes datos importantes: posiciones y tiempos.

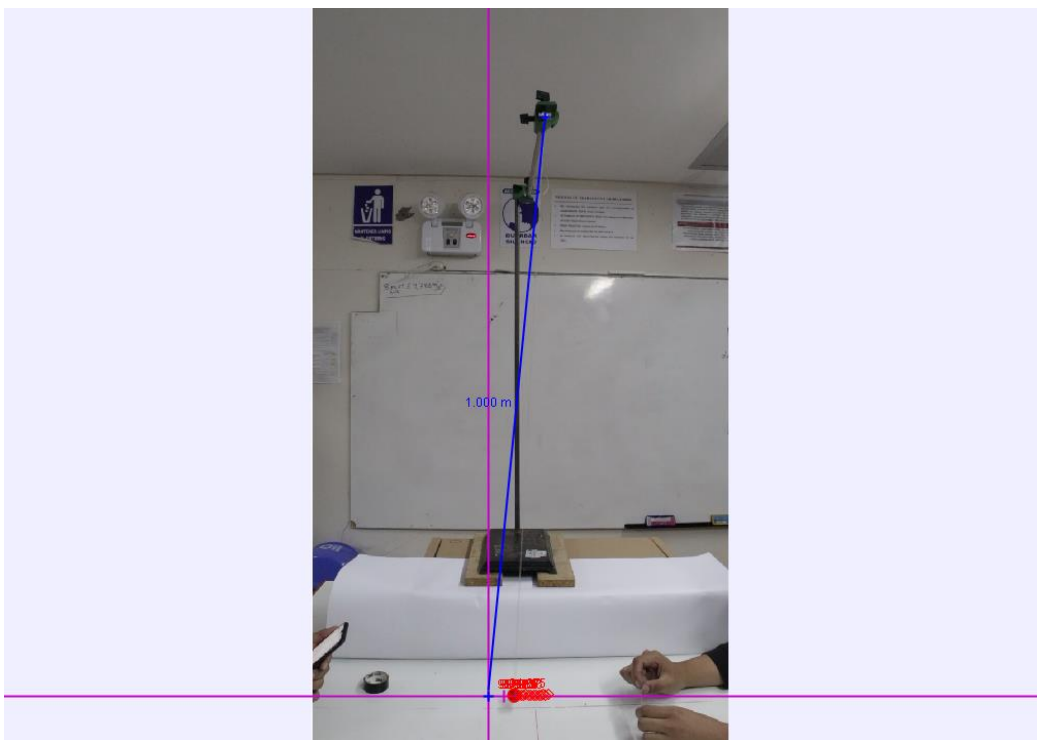

*Figura 6.* *Movimiento del péndulo simple para L=1 m*

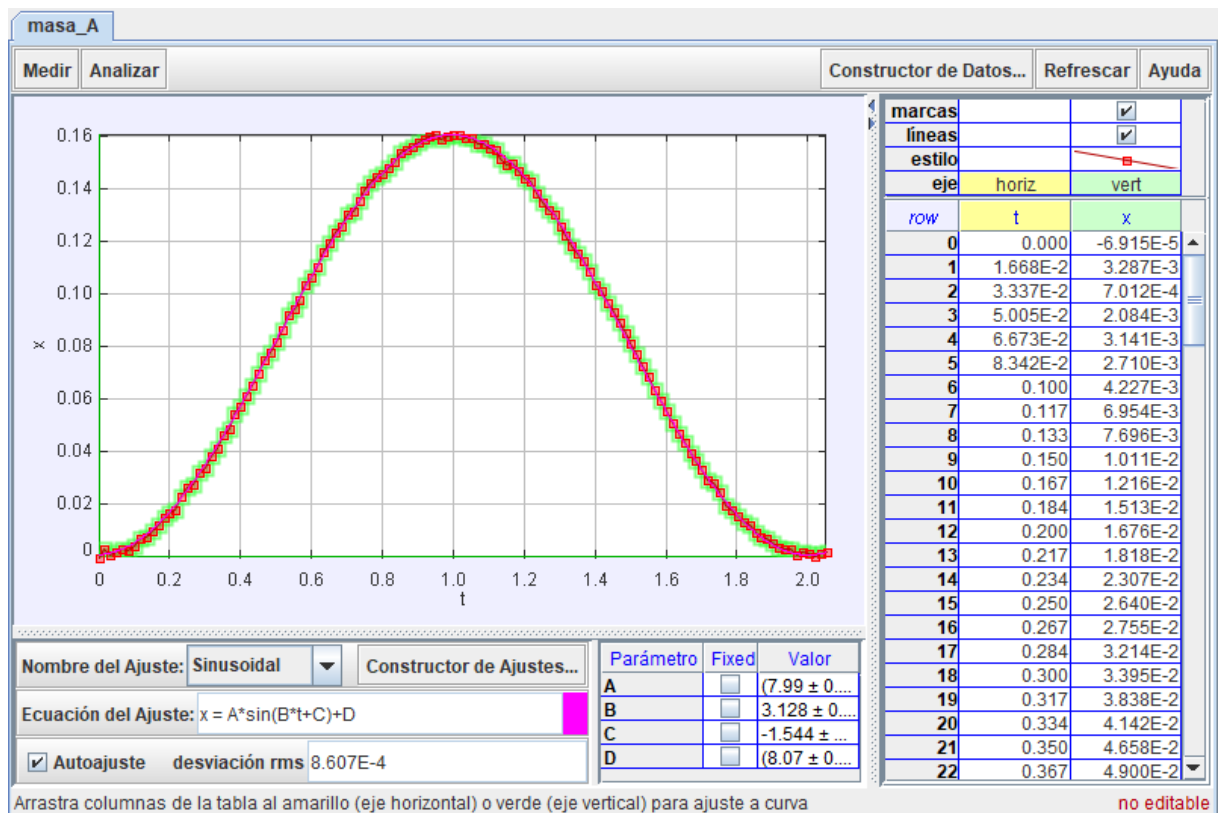

*Figura 7. Ecuación del ajuste y su gráfica obtenida por los datos tomados con el rastreo para L=1 m*

**Análisis de datos:**

El software Tracker ya nos brinda un previo conocimiento acerca de la curva y el formato que esta tiene, en ese sentido, se estableció que el ajuste de la curva debía de ser una curva sinusoidal con 4 parámetros por determinar, es decir:

$$x(t) = A\sin(Bt \pm C) + D \qquad (14)$$

**Parámetros a calcular:**

$A$: representa la amplitud de la ecuación de movimiento del péndulo simple.
$B = \omega$: representa la frecuencia angular del péndulo simple.
$C = \phi$: representa el ángulo de desfase del péndulo simple.
$D$: representa la proyección inicial del péndulo simple.

**Exportación de datos desde Excel a Python en Google Colab**

Para exportar los datos a Python, usaremos Google Colab que nos permite usar Python en línea, donde primero, copiamos los datos del Tracker con total precisión, es decir con todos los decimales a Excel para después guardar con el nombre de pila del proyecto.

Para subir el archivo colocamos el siguiente código en el Google Colab:

```python
# Subida de datos
import pandas as pd
import seaborn as sns

dfh_a = pd.read_excel("L=1m.xlsx")
dfh_a
```

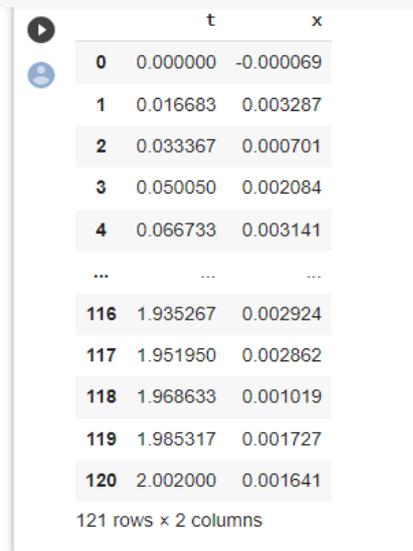

*Figura 8. Datos para "L=1m" obtenidos por el software Tracker.*

Se procede con la importación de librerías base para el análisis de datos.

```python
# Importación de librerías
import numpy as np
import sympy as sp
import matplotlib.pyplot as plt
from scipy.optimize import curve_fit
from pandas import read_csv
import pandas as pd
import seaborn as sns
```

Para la subida de datos y el correspondiente ajuste de curva:

```python
# Subida de datos
dfh_a = pd.read_excel("L=1m.xlsx")
dfh_a
# En este caso se declara la variable "data" para que los datos previamente exportados
sean transformados en un arreglo, esto con el objetivo de facilitar de mejor manera el
gráfico y ajuste de curva.
data = dfh_a.values
# Ahora, se declara 2 variables "x" e "y" en los cuales se incluyen los datos
previamente exportados, pero en forma de arreglo, para "x" se encuentran todos los
elementos del array de la columna 0 (valores del tiempo) y para "y" se encuentran todos
los elementos del array de la columna 1 (valores de la posición).
x = data[:,0]
y = data[:,1]
# Se grafican los datos de "x" e "y"
plt.scatter(x, y)
# Se declara una función "objective" en el cual se construye la función objetivo con la
cual se ajusta la curva los datos obtenidos, donde se incluyen los parámetros de los cuales
debe conocerse su valor.
def objective(x, a, b, c, d):
  return a * np.sin(b*x + c) + d
# Se definen las variables popt y pcov en base a la función objetivo.
# pcov, corresponde al arreglo de covarianza y da una estimación del error en los
parámetros.
popt, pcov = curve_fit(objective, x, y)
# Permite obtener los parámetros óptimos que definirían la curva.
a, b, c, d = popt
# Se usa linspace debido a que en este caso se debe definir los datos mediante secuencias
numéricas.
x_model = np.linspace(min(x), max(x), 100)
# Se define la curva de ajuste en base a los datos de x y los parámetros.
y_model = objective(x_model, a, b, c, d)
# Finalmente se procede con el gráfico y se definen las características de la curva de
ajuste.
plt.scatter(x,y)
plt.plot(x_model,y_model, '--', color='r')
```

```
plt.text(-0.03,    -0.03,    r'$X(t)='+str(a)+'sen('+str(b)+'t    +
'+str(c)+') + '+str(d)+' $', fontsize=10.5)
# Se agrega 2 cuadros de texto donde en uno se incluirá la curva ajustada con sus
respectivos parámetros calculados, y el otro incluirá una leyenda que clasificará los datos
experimentales con la curva ajustada a mostrarse, finalmente se procede a mostrar el
gráfico.
plt.legend(['Datos Experimentales', 'Curva Ajustada'])
plt.show()
```

- De los datos obtenidos, se obtuvo una gráfica sinusoidal, cuya ecuación de ajuste está dado por:

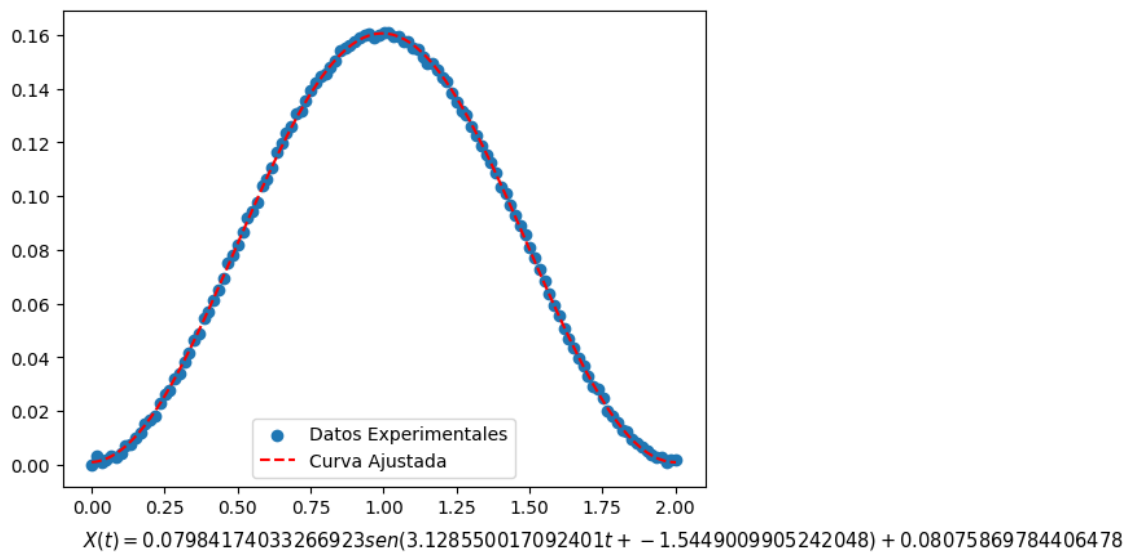

$X(t) = 0.07984174033266923 sen(3.128550017092401t + -1.5449009905242048) + 0.08075869784406478$

***Figura 9.*** *Gráfica sinusoidal obtenida en función a los datos experimentales y la curva ajustada*

Teniendo que la ecuación de la curva ajustada es:

$$x(t) = 0.0798417 sen(3.1285500t - 1.5449010) + 0.0807587 \quad (15)$$

Reconociendo parámetros:

$$A = 0.0798 \text{ m}$$
$$\omega = 3.1286 \text{ rad/s}$$
$$\phi = -1.5449 \text{ rad}$$
$$D = 0.0807 \text{ m}$$

Para el cálculo de la gravedad en función a la ecuación (10), se tiene

$$g = \omega^2 L \rightarrow g = (3.1286 \text{ rad/s})^2 \ (1.00 \text{ m})$$

$$g = 9.7878 \text{ m/s}^2$$

Realizamos el mismo procedimiento para el resto de longitudes del péndulo simple, obteniendo de esta manera los siguientes resultados:

*Tabla 3.*
*Valores de la aceleración de la gravedad obtenidos mediante el análisis de datos proporcionados por el software Tracker.*

| $L$ (m) | $\omega$ (rad/s) | $g$ (m/s$^2$) |
| --- | --- | --- |
| 1.00 | 3.1286 | 9.7878 |
| 0.95 | 3.1918 | 9.6781 |
| 0.90 | -3.2874 | 9.7263 |
| 0.85 | -3.3999 | 9.8255 |
| 0.80 | 3.5237 | 9.9333 |
| 0.75 | 3.6568 | 10.0291 |
| 0.70 | -3.6646 | 9.4005 |
| 0.65 | 3.8949 | 9.8607 |
| 0.60 | 4.0117 | 9.6562 |
| 0.55 | 4.1740 | 9.5823 |
| 0.50 | 4.3935 | 9.6514 |
|  | $\bar{g}$ | 9.7392 |

Fuente: Elaboración propia.

En función de los datos obtenidos por el método del Tracker podemos ver que tenemos una media de 9.7392 m/s² con una incertidumbre de 0.1744 m/s²

**3. Resultados**

Para evaluar los datos obtenidos, analizamos el error porcentual con la ecuación (11), para el valor teórico de la gravedad en función al valor de la gravedad obtenida por el péndulo simple y para el valor de la gravedad obtenido mediante el software Tracker.

$$E_{Péndulo\ Simple}\% = \frac{|9.7836 - 10.1760|}{9.7836} \times 100\% = 4.0108\ \%$$

$$E_{Tracker}\% = \frac{|9.7836 - 9.7392|}{9.7836} \times 100\% = 0.4538\ \%$$

Teniendo de esta manera los valores de la aceleración de la gravedad en la ciudad del Cusco:

$$g_{teórico} = 9.7836\ \text{m/s}^2$$

$$g_{Péndulo\ Simple} = 10.1760 \pm 0.1961\ con\ E\% = 4.0108\ \%$$

$$g_{Tracker} = 9.7392 \pm 0.1744\ con\ E\% = 0.4538\ \%$$

**4. Discusión y Conclusiones**

En esta investigación nos enfocamos en el análisis y evaluación de tres métodos utilizados para determinar la gravedad teórica. Los resultados obtenidos a través de estos métodos proporcionan información valiosa sobre la exactitud y confiabilidad de los mismos, así como su aplicabilidad en el contexto de la educación a distancia.

En primer lugar, se utilizó la ley de la gravitación universal de Newton para calcular la gravedad teórica en el Cusco, sin embargo, por las variables de la ecuación podemos permitirnos realizar modificaciones en función a cualquier ciudad en la que estemos situados. Al aplicar esta ley, se obtuvo un valor teórico de la aceleración de la gravedad en la ciudad del Cusco de 9.7836 m/s$^2$. Este resultado concuerda con investigaciones previas realizadas por Sebastiá en 2013, lo cual respalda la validez de la ley de gravitación universal como herramienta para calcular la aceleración de la gravedad en diferentes ubicaciones geográficas.

El segundo método empleado fue el método tradicional para la recolección de datos del experimento del péndulo simple, este método es generalmente usado en la educación básica y universitaria en laboratorios presenciales para poder determinar la aceleración de la gravedad de manera práctica. En nuestro caso seguimos las directrices usadas en los laboratorios universitarios, de esta manera se armó un péndulo utilizando una varilla de aluminio y una cuerda con longitud variable. Se midió el tiempo en el que el péndulo realiza 10 oscilaciones para cada variación de la longitud de cuerda y se utilizó la fórmula $x = L\sin(5°)$ para calcular el desplazamiento máximo sobre el eje x del péndulo. Se realizaron mediciones para longitudes de cuerda desde 1 metro, variando en 5 cm hasta 50 centímetros, y se registraron los datos obtenidos en la tabla 1.

A partir de estos datos, se aplicaron técnicas de análisis estadístico para obtener el valor de la gravedad utilizando el péndulo simple. El resultado obtenido fue de 10.1760 m/s$^2$, con una incertidumbre de 0.1961 m/s$^2$ con un error porcentual de 4.0108% respecto al valor teórico. Estos valores indican que el método del péndulo simple proporciona una estimación bastante cercana al valor teórico de la aceleración gravedad, aunque con un margen de error aceptable.

El tercer método utilizado fue el software Tracker, el cual se empleó para recolectar y analizar los datos del péndulo simple obtenidos mediante una filmación de cada variación en el mismo experimento, teniendo así un total de 11 filmaciones. A través del software Tracker, se obtuvo un valor promedio de la aceleración de la gravedad de 9.7392 m/s$^2$, con una incertidumbre de 0.1744 m/s$^2$ y un error porcentual del 0.4538% respecto al valor teórico. Estos resultados indican una mejor aproximación al valor teórico de la gravedad y una reducción significativa en el margen de error en comparación con el método del péndulo simple.

La importancia de estos métodos y herramientas computacionales para el contexto de la educación a distancia, de los cuales podemos destacar que en contextos como la pandemia de COVID-19 en la cual se ha adoptado la educación en línea, generó la necesidad de desarrollar estrategias y herramientas efectivas para la enseñanza de asignaturas prácticas, como la física. En este sentido, tanto el péndulo simple como el software Tracker ofrecen soluciones viables para realizar experimentos y análisis de datos de manera remota.

Validamos mediante esta investigación que el software Tracker brinda la posibilidad de realizar recolección y análisis de datos en cualquier lugar, donde los estudiantes pueden analizar el movimiento de objetos y realizar mediciones precisas utilizando videos grabados previamente.

En conclusión, se observa que tanto el método tradicional de recolección de datos del péndulo simple como el uso del software Tracker proporcionaron valores cercanos al valor teórico de la aceleración de la gravedad en el Cusco. Sin embargo, el método del software Tracker presentó un menor margen de error, lo que indica una mayor precisión en las mediciones y cálculos realizados. Esto evidencia la importancia de utilizar herramientas tecnológicas en la educación a distancia, ya que permiten obtener resultados más confiables y consistentes.

**Agradecimientos**

**Referencias**